# Gleason Score Prediction using Deep Learning in Tissue Microarray Image

Yi-hong Zhang[a, #], Jing Zhang[a, #], Yang Song[a], Chaomin Shen[b], Guang Yang[a, *]

a, Shanghai Key Laboratory of Magnetic Resonance, East China Normal University, Shanghai, China

b, School of Computer Science, East China Normal University, Shanghai, China

\# These authors contributed equally to this work.

*Corresponding Author: Guang Yang, gyang@phy.ecnu.edu.cn. Tel: +86-021-62233873

**Abstract:** Prostate cancer (PCa) is one of the most common cancers in men around the world. The most accurate method to evaluate lesion levels of PCa is the microscopic inspection of stained biopsy tissue and estimate the Gleason score of tissue microarray (TMA) image by expert pathologists. However, it is time-consuming for pathologists to identify the cellular and glandular patterns for Gleason grading in large TMA images. We used Gleason2019 Challenge dataset to build a convolutional neural network (CNN) model to segment TMA images to regions of different Gleason grades and predict the Gleason score according to the grading segmentation. We used a pre-trained model of prostate segmentation to increase the accuracy of the Gleason grade segmentation. The model achieved a mean Dice of 75.6% on the test cohort and ranked 4[th] in the Gleason2019 Challenge with a score of 0.778 combined of Cohen's kappa and the f1-score.

# 1 Introduction

Prostate cancer (PCa) is the second common cancer and also one of the most leading causes of cancer death among men worldwide [1]. One of the most accurate methods to detect PCa is the microscopic inspection of stained biopsy by a pathologist. Regions of tissue are assessed and given a Gleason grade of 1 to 5 according to the observed histological patterns. The sum of the most prominent and second most prominent patterns is the final Gleason score. The assessment of Gleason score is not only time-consuming, but also relies on the experience of the pathologist and suffers from very high inter-observer variability.

The computer-aided diagnosis method based on convolutional neural network (CNN) has played an important role in medical image segmentation and detection [2, 3]. So, we used CNN to identify different Gleason grade regions in the tissue micro-array (TMA) images of PCa and report the final Gleason score.

# 2 Methods and Materials

We used TMA images from the Gleason Challenge 2019 in this study [4]. The sizes of the images ranged from 4608 to 5632. Each TMA image was annotated by six professional pathologists independently (Figure 1). Each pixel was marked as one of benign, Gleason grade lower than 3, Gleason grade 3, 4, 5, and ignored region. We split 244 cases into the training (188), validation (33) and test (23) cohorts. The details of the distribution of annotation were shown in Table 1.

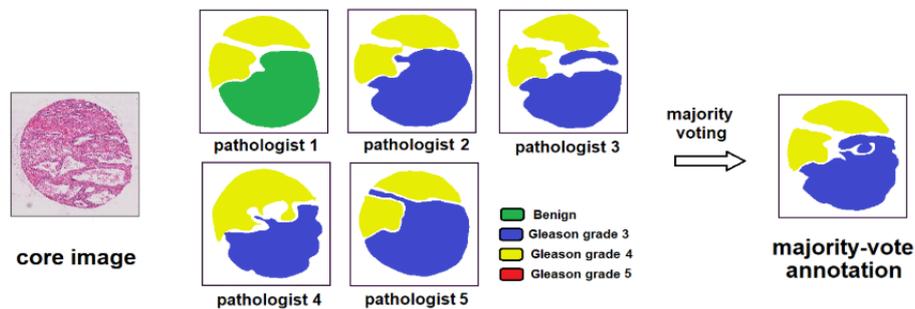

Figure 1. images annotated by six professional pathologists[4].

Table 1. The distribution of Gleason grade in the training, validation and test cohorts

|  | Total cases | Benign | G=3 | G=4 | G=5 |
|---|---|---|---|---|---|
| Train | 188 | 72 | 111 | 134 | 10 |
| Validation | 33 | 13 | 20 | 23 | 1 |
| Test | 23 | 15 | 10 | 14 | 3 |
| Total | 244 | 100 | 141 | 171 | 14 |

To reduce the complexity of the model, we down-sampled the TAM images by a factor of 10, using B-spline interpolation. Then, we cropped or zero-padded the images to a fixed size of 448×448. Then we normalized the images by Z-score standardization through all images in the cohort.

The corresponding annotations were annotated by six pathologists. We first encoded the annotated values onto six channels, respectively. Then we merged annotations by different pathologists for each channel to reduce the inter-variance among these annotations (Figure 2). To increase the robustness of the model, we augmented training cases by randomly flipping, rotating, or stretching. Finally, the sizes of the input and output are 448×488×3 and 448×448×6, respectively.

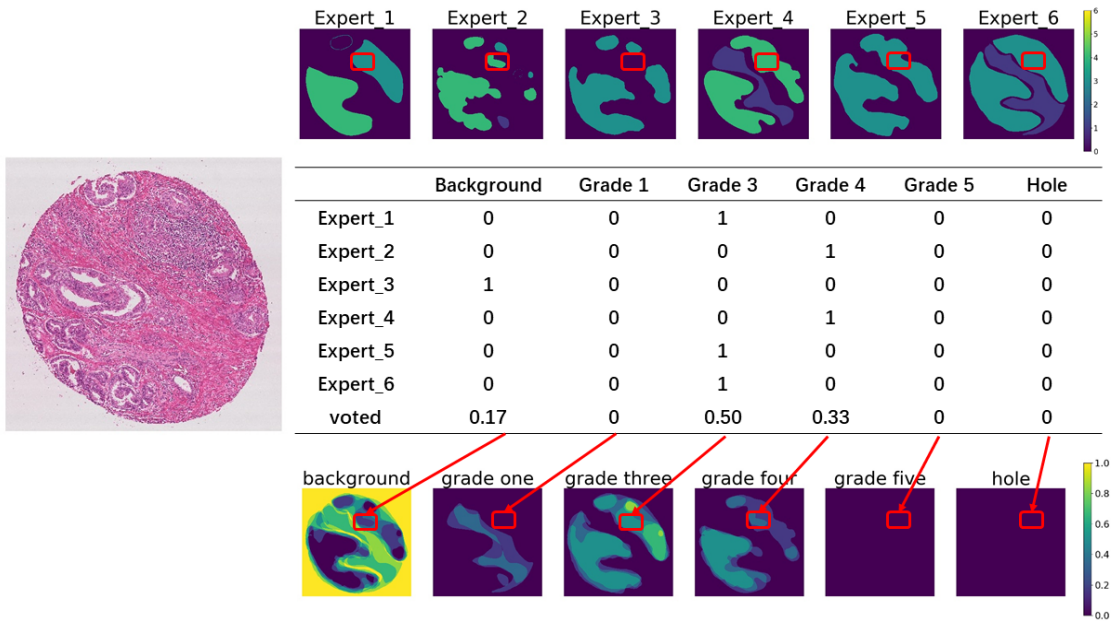

Figure 2. The preprocessing of the images. Each annotation by the pathologists were encoded into 6 channels. Then we merged annotations through the channels and got the expected probability map for each channel.

We used a network based on U-Net [5] to segment a TMA image into different regions (Figure 3). The network included two parts: encoding part and decoding part. Encoding part contained four convolutional blocks, and each block had two convolutional layers of kernel size 3×3, followed by Batch Normalization [6] and ELU activation layer [7]. We applied the max-pooling between two convolutional blocks to increase the size of receptive field and decrease the size of the feature map. We used 64, 128, 256, and 512 filters in these four blocks. In decoding part, three convolutional blocks with 256, 128, and 64 filters were used to decode the feature map into the size of inputs. We deconvoluted feature maps and concatenated the corresponding features by skip connection. Finally, we applied a convolutional layer with six 1×1 followed by a softmax activation to get the probability map of Gleason grade. We used Adam [8] to optimize the network, and the cross-entropy as the loss function. We implemented all above processes based on Python 3.6 with TensorFlow 1.13.1 and Keras 2.2.4

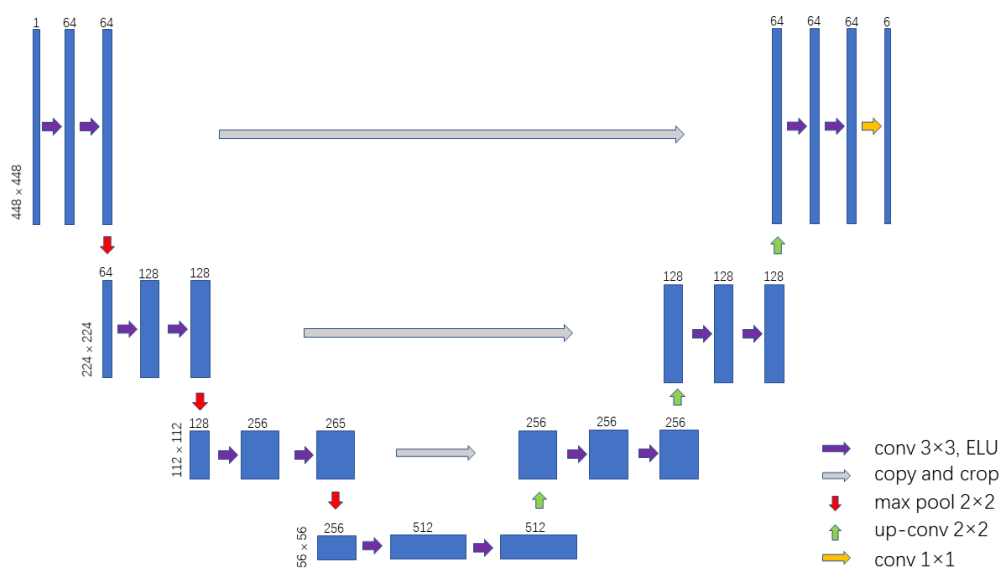

Figure 3. The architecture of the U-Net based network. Each blue box denotes the feature map with multi-channels.

We merged the probability map through the channel direction and got the final region segmentation. Then we used a median filter to remove the noise and the indeterminate value from the prediction map. The size of the filter was chosen to 55×55 based on the performance of the validation cohort. We up-sampled the segmentation to the raw size and got the final Gleason grade by the nearest neighbor interpolation. Finally, we estimated the area of most prominent and the second most prominent to predict the final Gleason score.

We also compared the model with initial-random parameters (Model 1) and the model with pre-trained parameters (Model 2) [9], which was trained by a prostate segmentation task with 150 T2 weighted images.

We used Dice coefficient to evaluate the performance of the segmentation. The model was evaluated on the independent testing cohort by the confusion matrix and quantitative score combined by Cohen's Kappa and the F1-scores. The final score was computed as follows.

$$Score = Cohen's\ Kappa + \frac{F_{1,macro} + F_{1,micro}}{2}$$

3 Results

The Dice of Model 1 on test cohort achieved an average of 0.749 and Model 2 achieved 0.756. We showed the Dice distribution of each images in Figure 4. We found that the mean Dice of Model 2 was higher than that of Model 1 by 0.007. In addition, the Dice of 91% of cases in Model 1 and 95% of Model 2 exceeded 0.6. We showed segmentations of three images randomly in Figure 5.

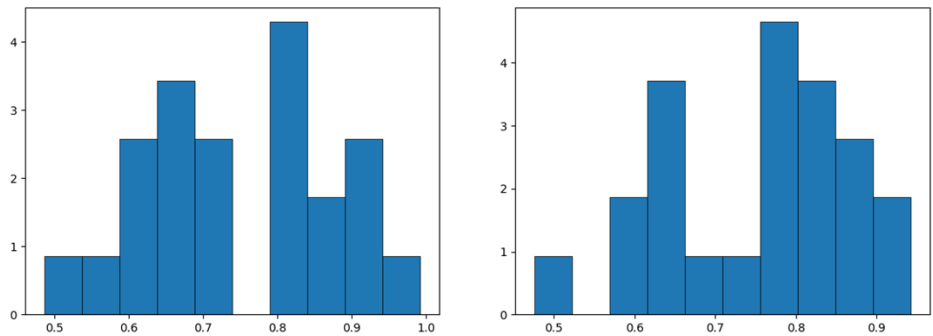

Figure 4 The Dice distribution of two models on the test cohort. The average of Dice achieved 0.749 for the Model 1 (left) and 0.756 for the Model 2 (right).

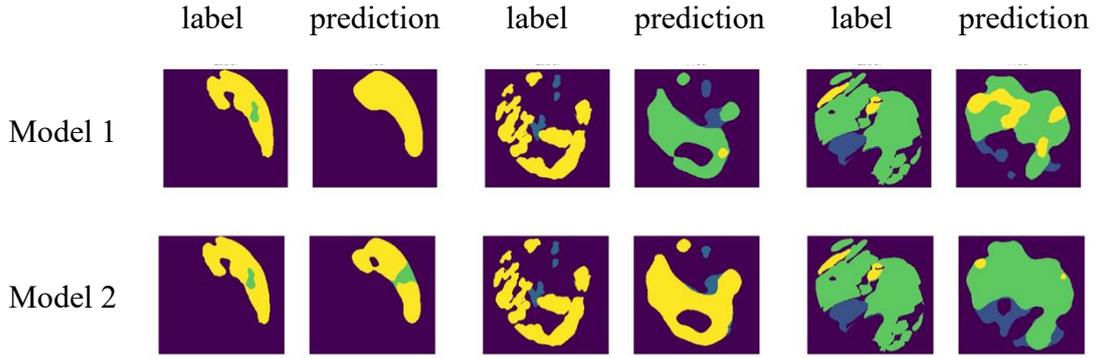

|   | label | prediction | label | prediction | label | prediction |
|---|---|---|---|---|---|---|
| Model 1 | | | | | | |
| Model 2 | | | | | | |

Figure 5 Three cases were randomly selected for visualization. The top row denotes the prediction of Model 1, and the bottom row denotes that of the Model 2. In each case, the annotation image is on the left and the prediction is on the right.

Model 2 achieved a score of 0.778 on the Gleason2019 Challenge and ranked the 4[th]. The confusion matrix was shown in the Result of Gleason2019 Challenge [4]. The classification accuracy of benign (88.32%) was higher than that of the Gleason grade of 3 (66.57%), while the accuracy of Gleason grade of 4 (46.93%) and 5 (22.90%) were much lower.

4 Discussions and Conclusions

We used a pre-trained U-Net based network to estimate the Gleason grade region from MTA images and the model achieved an average Dice of 0.756. The model ranked the 4[th] with a score of 0.778 on the Gleason2019 Challenge. We estimated the region marked as Gleason grade five lower than other regions, which may be related to the complex structure of the cells and the imbalanced cases. More cases with an annotation of high Gleason grade could help the network learn the structure to identify the region of high Gleason grades. Several pre-processing and post-processing were used to reduce the complexity and make the CNN model easy to estimate the different regions. However, these processes, such as interpolation and filtering, reduced the details of the raw TMA images and the prediction map, which leaded to a smooth segmentation but lacked an accurate estimation. Previous studies on the pathological images focused on the cell identification from the core images, but this requires a large amount of time on the annotation [10]. Semi-supervised and un-supervised learning are critical for TMA to extract the features and identify the Gleason grade [11]. Some training strategies and network designs, such as transfer learning, generative adversarial networks and multi-tasks, may help increase the performance of the model when using the small data set [12-14]. TMA analysis and Gleason grade identification took much time and there is inter-variance among different pathologists. Convolutional neural networks have powerful potential to reduce the variance and aid the pathologists diagnose the Gleason score in the clinics.

5 Acknowledgments